\documentclass[lettersize,journal]{IEEEtran}
\usepackage{amsmath,amsfonts}
\usepackage{algorithmic}
\usepackage{algorithm}
\usepackage{array}
\usepackage[caption=false,font=normalsize,labelfont=sf,textfont=sf]{subfig}
\usepackage{textcomp}
\usepackage{stfloats}
\usepackage{url}
\usepackage{verbatim}
\usepackage{graphicx}

\usepackage{tikz}
\usepgflibrary{plotmarks}

\usepackage{cite}
\usepackage{amssymb}
\usepackage{amsmath}
 
\DeclareMathOperator*{\argmax}{argmax} 
\usepackage{hyperref}       % hyperlinks

\usepackage{booktabs}       % professional-quality tables
\usepackage{cuted}
\newcommand{\commentall}[1]{}

\begin{document}

\title{Modelling the nanopore sequencing process with Helicase HMMs}

\author{Xuechun Xu~\IEEEmembership{Student member,~IEEE}, Joakim Jaldén~\IEEEmembership{Senior member,~IEEE}
        \thanks{X. Xu and J. Jal\'{e}n are with the division of Information Science and Engineering, KTH, Royal Institution of Technology, Stockholm, Sweden.}}% <-this % stops a space}

\markboth{Journal of \LaTeX\ Class Files,~Vol.~14, No.~8, August~2021}%
{Xuechun Xu, Joakim Jald\'{e}n: Marginalized Beam Search Algorithms for Hierarchical HMMs}


\maketitle

\begin{abstract}
Recent advancements in nanopore sequencing technology, particularly the R$10$ nanopore from Oxford Nanopore Technology, have necessitated the development of improved data processing methods to utilize their potential for more than $9$-mer resolution fully. The processing of the ion currents predominantly utilizes neural network-based methods known for their high basecalling accuracy but face developmental bottlenecks at higher resolutions. In light of this, we introduce the Helicase Hidden Markov Model (HHMM), a novel framework designed to incorporate the dynamics of the helicase motor protein alongside the nucleotide sequence during nanopore sequencing. This model supports the analysis of millions of distinct states, enhancing our understanding of raw ion currents and their alignment with nucleotide sequences. Our findings demonstrate the utility of HHMM not only as a potent visualization tool but also as an effective base for developing advanced basecalling algorithms. This approach offers a promising avenue for leveraging the full capabilities of emerging high-resolution nanopore sequencing technologies.
\end{abstract}

% Inspired by paper 'Analysis of nanopore data using hidden Markov models'
\section{Introduction}
Hidden Markov models (HMMs) are integral tools in the field of genome analysis, playing a crucial role across various stages of DNA exploration. These models offer a versatile framework for representing the DNA strands in various complexities, proving particularly useful in different tasks like sequencing, assemblies, and mapping. At the foundational level, HMMs efficiently capture the essence of the DNA strand through discrete state spaces, where the concept of $k$-mers, representing $k$ consecutive nucleotide bases, provides a measurable unit. In later stages of genome analysis, including genome notation, variance detection, and epigenetic studies, HMMs demonstrate their adaptability with the state spaces representing functional units of DNA, such as exons and introns. 

Although HMMs have been popular because of their flexibility and applicability throughout the genomic analysis pipeline, the landscape of genome analysis has witnessed a notable shift with the rapid evolution of (deep) neural networks (NNs), especially in the application of basecalling in (long read) nanopore sequencing. Nanopore sequencing technology utilizes nanometer-scale protein pores, i.e., a nanopore, to measure the blockage of multiple nucleotides of the DNA strand on an ion current, as a motor protein ratchets the DNA through the nanopore one base at a time. The resulting ion currents are measured and later transformed into nucleotide bases via an algorithm referred to as a basecaller.

During the early development of the nanopore sequencing technology, basecallers predominantly relied on HMMs. A segmentation step was used to separate the ion currents into relatively stationary events, also known as `squiggles', by non-supervised $t$-tests or a separate segmentation NN \cite{ontyoutube,baseRawller}. These events were crucial for basecalling/variant detection and were part of deliverable results by Oxford Nanopore Technology (ONT) in the initial versions. The HMM was tasked with modelling the segmented event \cite{ONTdelivery2016}. However, such a segmentation step also introduces errors that may subtly accumulate in subsequent modelling phases. The HMM-based tools \cite{analysiswithHMM,Nanocall,HMMHDP} thus either had to model the segmentation errors on top of modelling the already complicated sequencing process or performed a re-segmentation, also known as the `re-squiggle' \cite{Nanoraw}. Researchers must often introduce additional states to increase the model complexity, e.g., extra deletion/skip states corresponding to no emission states, as the segmentation could omit bases \cite{Nanocall,Nanopolish}.

In contrast, the newer NN-based methods sidestep segmentation errors by directly processing raw ion currents and predicting the nucleotide sequence \textit{end-to-end}. A pioneering study of end-to-end basecalling, which led to the Chiron basecaller \cite{Chiron}, first demonstrated the usage of a Connectionist Temporal Classification (CTC) decoder on top of an NN for this application, which, in a sense, replaced the segmentation step. Since then, researchers have diligently refined NN-based basecallers with the CTC layer or a similar structured Conditional Random Field (CRF), solidifying their superiority in accuracy over their HMM-based counterparts. Nowadays, both the state-of-the-art commercial and research software published by ONT are NN-based frameworks. %The application of NNs in the context of Nanopore sequencing technologies has garnered attention for their superior sequencing accuracy compared to HMM-based methods.

%\begin{figure}[!tp]
% \centering
%  \includegraphics[width=0.9\columnwidth]{figure/variance_basepos.png}
%  \caption{Average variance of the ion currents for one base difference on a $k$-mer. The values change if the varying base appears in different positions. The blue line displays the extracted mean from the $11$-mer Helicase HMM observations, and the orange is from the ONT $9$-mer table for the R10 nanopore. } % JJ: We need to expand this with a more detailed explanation of what is shown in this picture. As it is now, this is not clear from the text. Can we redraw this figure? I would suggest using the method (I used before) that can yield a much longer influence dependency plot over many more positions.
%  \label{kmer}
%\end{figure}

While NNs present advantages in terms of sequencing accuracy, they introduce challenges and are at risk of facing a bottleneck. One such challenge is the model mismatch between the NN-based basecaller and other HMM-based tools utilized in subsequent genome analysis stages, which may cause additional costs of training different models on the same data set. One example is the recently achieved milestone study of the first gap-free human genome \cite{2022human,epigenetic2022}, where researchers leveraged the HMM-based tool Nanopolish \cite{Nanopolish} for insightful epigenetic studies despite using the NN-based basecaller. This strategy demands distinct training of the Nanopolish model on the raw ion currents due to the asymmetry in knowledge transferability between NN outputs and the state space of Nanopolish. The time- and energy-intensive retraining process could be circumvented by employing an HMM-based structure in the basecaller. In a previous study, we demonstrated, with a novel basecaller called Lokatt, that one could replace the CTC/CRF decoder with an arbitrarily designed HMM decoder to increase the adaptivity \cite{Lokatt} while maintaining high basecalling accuracy. 

\begin{figure}
\begin{center}
\begin{tikzpicture} [x=0.25cm,y=2.5cm]

\draw[help lines, ystep=0.2, xstep=1] (-15,-1) grid (15,1);
\draw[fill, color=gray, opacity=0.15] (-15,-1) rectangle (-7.5,1);
\draw[fill, color=gray, opacity=0.15] (2.5,-1) rectangle (15,1);
\draw (-15,-1) rectangle (15,1);
\foreach \x in {-15,-12,...,15} {
	\draw[thick] (\x,-1)+(0,0.5mm) -- +(0,-1.0mm) node [anchor=north] {\scriptsize $\x$};
}
\foreach \y in {-1,0,...,1} {
	\draw[thick] (-15,\y)+(-0.5mm,0) -- +(0,1.0mm) node [anchor=east] {\scriptsize $\y$};
}
\draw[mark=o] plot file{figure/ConditionalSignalValueA.txt};
\draw[mark=square] plot file{figure/ConditionalSignalValueC.txt};
\draw[mark=triangle] plot file{figure/ConditionalSignalValueG.txt};
\draw[mark=diamond] plot file{figure/ConditionalSignalValueT.txt};

\draw[fill, color=white] (11,0.1) rectangle (14,0.9);
\draw[] (11,0.1) rectangle (14,0.9);
\draw plot[only marks,mark=o] coordinates{(12,0.8)} node [right] {A};
\draw plot[only marks,mark=square] coordinates{(12,0.6)} node [right] {C};
\draw plot[only marks,mark=triangle] coordinates{(12,0.4)} node [right] {G};
\draw plot[only marks,mark=diamond] coordinates{(12,0.2)} node [right] {T};

\node at (0,-1.18) [anchor=north] {Base position relative to alignment};
\node at (-16.5,0) [anchor=south, rotate=90] {Conditional mean};

\end{tikzpicture}
\end{center}
\caption{Mean (normalized) R10 ONT raw signal value conditioned on individual nucleotide base at a given (relative) position. Over a large set of sequences aligned to their respective raw signals, the value on the y-axis is calculated by computing the average signal value overall signal values where a base at a position relative to the aligned base attains the value indicated by the legend. The lighter area indicates the sequence of bases that forms the $10$-mers used in this work. Our $10$-mers extend the $9$-mers used in ONTs Remora toolkit by the leftmost base at relative position $-7$, which allows us to more effectively handle the step-back mechanism described later.}
\label{kmer}
\end{figure}
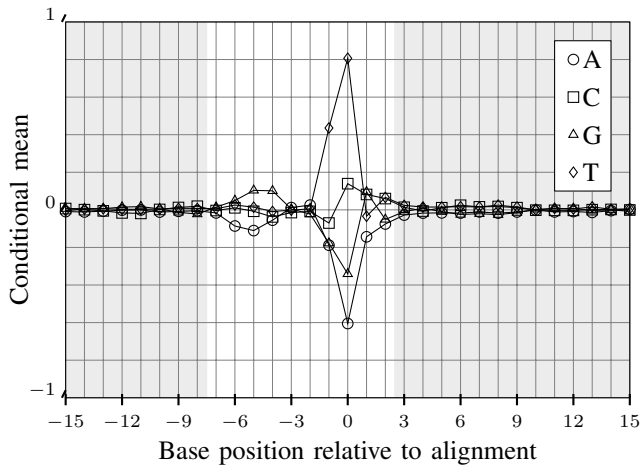

That said, Lokatt and other $ k$-mer-based NN models face a bottleneck where they are infeasible to definitely model the novel nanopore development with longer measurement depth, i.e., a state space representing $k$-mer with a relatively large $k$. As Figure~\ref{kmer} shows, the ion current measured by the latest R10 ONT nanopore is non-negligibly influenced by at least $9$ consecutive nucleotides, longer than the $5$-mer or $6$-mer of the previous R9 nanopore.
%JJ: We are mixing 9-mer and 10-mer in a way that makes this text confusing. We need to add more words here to explain exactly what we mean. Can we make the switch from talking about 9-mers to 10-mers in a more elegant way? I suggest a change below.
This $9$-mer resolution increases sequencing resolutions and substantially benefits the basecalling accuracy \cite{ONTR10}, in particular when calling homopolymers. However, given the number of classes to classify per measurement, a much larger NN output space is also necessitated under the standard mixed NN and HMM paradigm. If each $9$-mer is given its own class, the number of NN outputs is $4^{9}=262,144$. This poses impossible computational challenges surpassing the capacities of modern GPUs. To bypass the challenge, the basecallers trained for the R10 nanopore have to compromise and remain in the same state space as used for the R9 pores, e.g., ONT's cutting-edge research software Bonito uses a $5$-mer space in the `superior accuracy' model and a $3$-mer in the `fast' model \cite{Bonito}. 

In this paper, we re-explore and expand the potential of HMMs to address these limitations of NNs. In particular, we introduce a hierarchical HMM model seamlessly integrating insights into the behaviour of the helicase motor, including a step-back mechanism. We call this model the Helicase HMM.\footnote{Since the exact enzyme used as the motor can vary, we use `helicase motor' to refer to the motor protein's unzipping helicase-like function.} The modelled mechanism of the helicase motor was inspired by two elegant studies: In \cite{helicasedynamic}, the author observed the HEL308 DNA helicase through the high-resolution nanopore tweezers and reported the HEL308 pulling a nucleotide base half-back through the pore. In the study of an alternative motor enzyme, the $\phi29$ DNA polymer \cite{analysiswithHMM}, a similar step-back feature was also introduced in an HMM as a heuristic solution for efficiently correcting segmentation errors.

Modelling the $10$-mer with our Helicase HMM requires fewer parameters compared with the NNs. However, optimal implementation of the resulting base inference, i.e., the basecalling problem, is still infeasible. For example, we can no longer use the popular Viterbi algorithm as it is computationally infeasible to involve all $4^{10}$ conditional emission probabilities for each measurement/event. That said, we use the adapted Beam Search (BS) algorithm to trim the large search space and find approximate decoding solutions. Specifically, we use our novel Marginalized BS algorithm \cite{MBS} to solve the basecalling problem. 

In the subsequent sections, we introduce our proposed large-scale Helicase HMM model and algorithms meticulously tailored to navigate this expansive state space. To demonstrate the utility of the model, we validate our proposal in the experiment section, first with the re-squiggling task on analyzing ion current signals and second with the basecalling task, both utilizing human genome data sequenced by ONT R10 material. With the Helicase HMM, we wish to shed light on the HMM framework again. 

\section{Model structure}
The traditional HMM is a generative model used to characterise the nanopore sequencing process of a DNA sequence and the resulting ion current measurements. To reflect that the measuring region of the nanopore is stretched $k$ bases long, the reference DNA sequence is rewritten as a sequence of $k$-mers where each consecutive pair have $k-1$ overlapping bases. For example, a sequence `CAATAG' can be rewritten as a $3$-mer sequence: `CAA, AAT, ATA, TAG.' In the HMM framework, these $k$-mers are used as the hidden states of the sequence. The $k$-mer sequence is herein denoted by $\mathbf{K}=(K_1,K_2,...,K_M)$, with $K_m\in\mathbb{K}=\{1,2,...,4^k\}$, while the corresponding ion current measurements of the observation sequence are denoted by $\mathbf{X}=(X_1,X_2,...,X_N)$. Typically, the $k$-mer sequence is shorter than the current sequence, i.e., $M < N$, as the motor enzyme ratchets the DNA strand through the nanopore during the sequencing process and generates multiple ion current measurements per $k$-mer. 

That being said, it is very challenging for the basic HMMs to resemble the full sequencing process, which includes effects such as e.g. the motor proteins behaving differently as the chemical environment changes. Thus, to capture these complex yet unique effects, we extended the capability of basic HMMs and introduced the novel helicase HMM, particularly inspired by the motor-protein model. Our novel model distinguishes itself firstly from traditional HMM-based counterparts by modeling the $k$-mer sequences and ion current sequences end-to-end, which is achieved by incorporating a two-level hierarchical state space structure. Secondly, the Helicase HMM can handle classification over a massive state space compared to its NN-based counterparts. For reference, the $10$-mer states that we will later use, combined with a 5-state inner state model introduced shortly, results in a state space with over five million states.

\begin{figure}[tp]
  \centering
  \includegraphics[width=0.9\columnwidth]{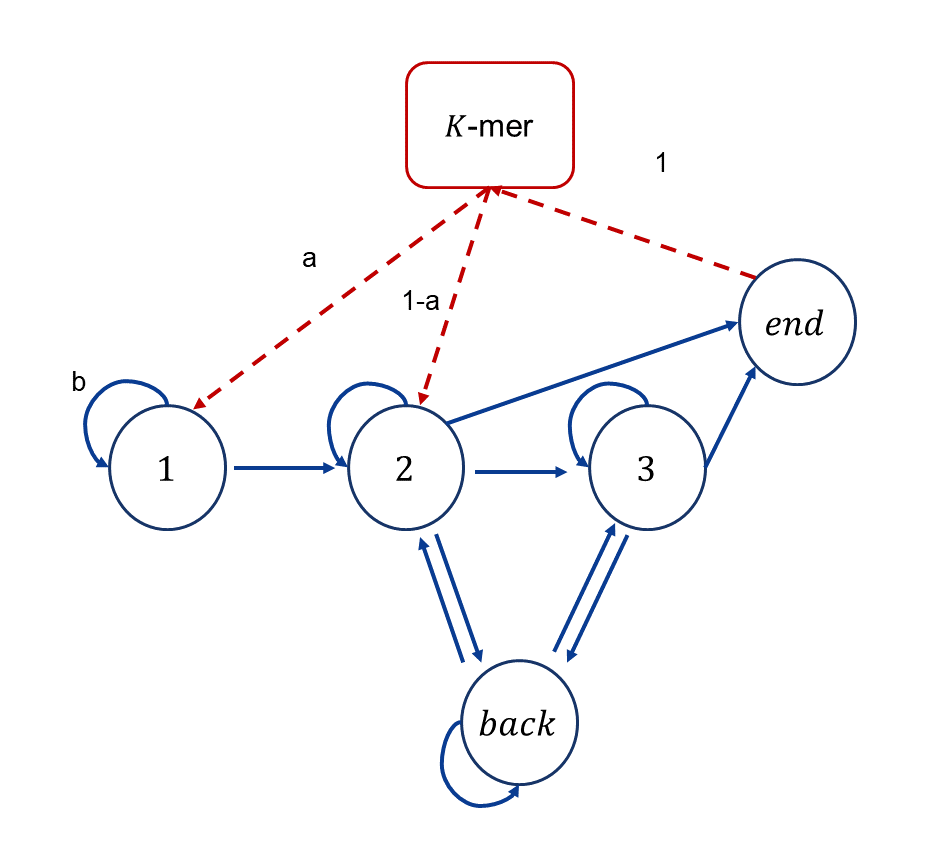}
  \caption{Illustration of the Helicase HMM and features of the nanopore reads. The two-level hierarchical structure of the Helicase HMM state space. On the top is the outer state representing $k$-mer, and at the lower are the $5$ inner states. The two dashed solid arrows pointing downward indicate the vertical transition that activates the inner states at the beginning of the domination of the corresponding outer state. The red dashed arrow pointing upwards indicates the vertical transition when the end inner state is activated and forces the domination to end. The blue solid arrows indicate the allowed transitions between all the inner states. The number above all arrows is the probability of such a transition.}
  % JJ: Should we mark the probabilities used for all arrows?
  \label{HelicaseHMM}
\end{figure}
\subsection{Helicase HMM}
% Basic HHMM concept, state, observation, how it aligned
The efficacy of the Helicase HMM in modeling the sequences end-to-end stems from its fundamental two-order hierarchical state space structure. As depicted in Figure~\ref{HelicaseHMM}, at the first level, the Helicase HMM has a $k$-mer state $K\in\mathbb{K}$, that we refer to as the outer state, in contrast to the inner states at the lower level, denoted by $(H_{(m,1)},H_{(m,2)},...,H_{(m,L_m)})$ under the $m$th outer state. Note that the inner state is the emission state, and $L_m$ is hence the duration, i.e., the number of measurements, of the $m$th $k$-mer. These newly introduced inner states represent the underlying dynamic of the sequencing process and observations. Thus, in the two-level hierarchy, the inner states are dominated by the outer states, such that the dominating outer state activates the chain of the inner state. Still, their horizontal transition does not affect the outer state until it reaches the end. 

\begin{figure*}[!htp]
  \centering
  \includegraphics[width=0.9\textwidth]{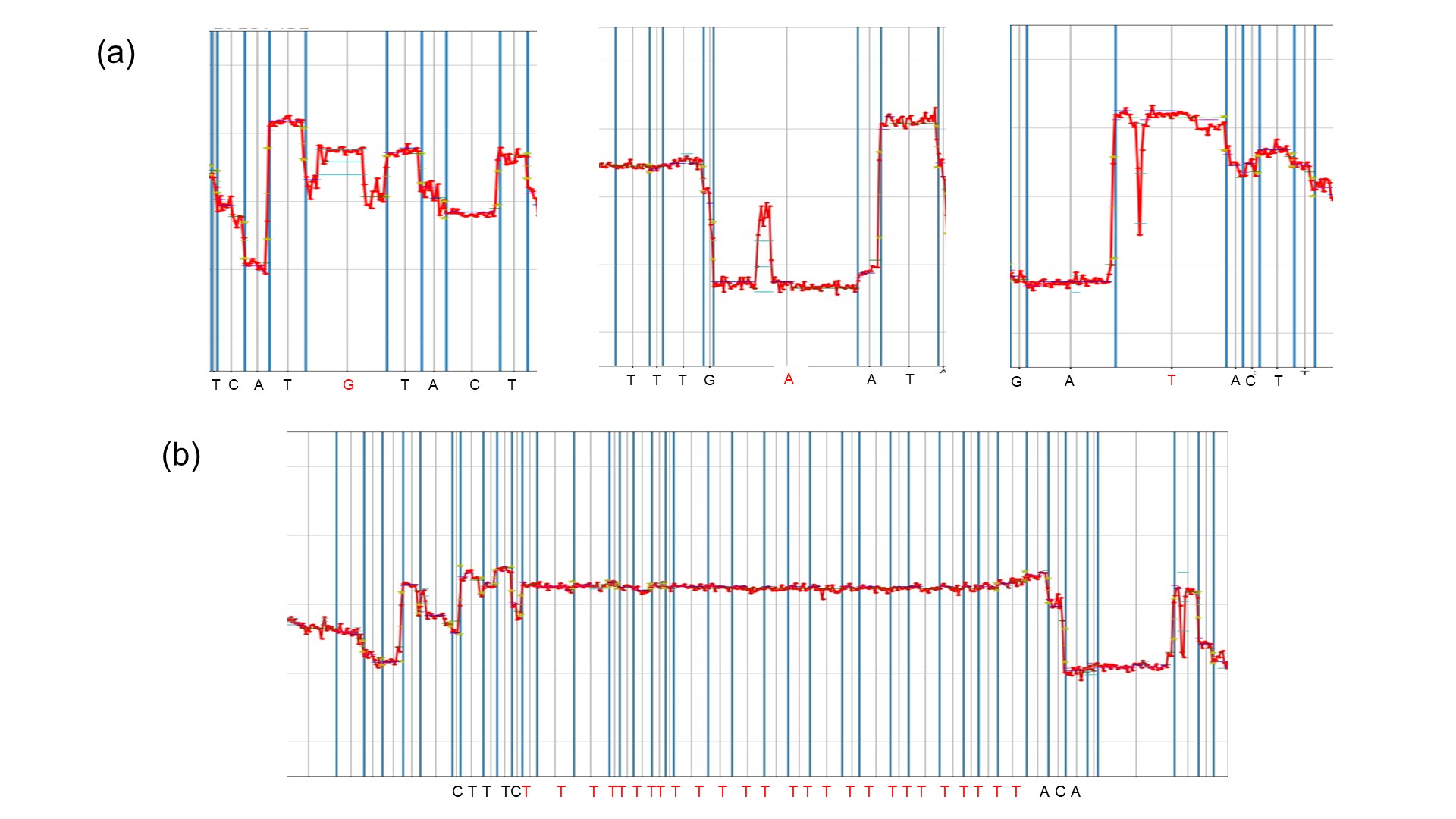}
  \caption{R10 nanopore ion currents with alignments (vertical blue lines) computed by re-squiggling with the Helicase HMM. The $7$th base of the $10$-mer is plotted as the major factor at the bottom. (a) Three examples of the step-back feature of the raw nanopore ion currents. The $10$-mers estimated to experience the step-back are marked by the red color. Within their dominating duration, the signals fluctuate towards the level of the previous $10$-mer. (b) An example of the alignment over the homopolymers region. The signal values are almost identical over $22$ bases of T.}
  % JJ: Need to clarify how the alignments were done
  \label{signals}
\end{figure*}

The inner state space is structured to comprise an `end' state and four helicase states, namely ${1,2,3}$ and `back', i.e., $H\in\mathbb{H}=\{end,1,2,3,back\}$. Notably, the inclusion of these four helicase inner states is informed by pioneer studies: Three observable states and the half-state back step were also suggested in \cite{helicasedynamic}, while similar transition mechanisms, `rapid flicker' and `backslips,' were introduced in \cite{analysiswithHMM}. These findings were further confirmed in the practical data used in this work. In Fig.~\ref{signals}(a), we provide three graphical evidence of such `step back' mechanisms, where the ion currents fluctuate towards the measured values of the preceding bases.

\begin{figure}[t]
 \centering
  \includegraphics[width=0.9\columnwidth]{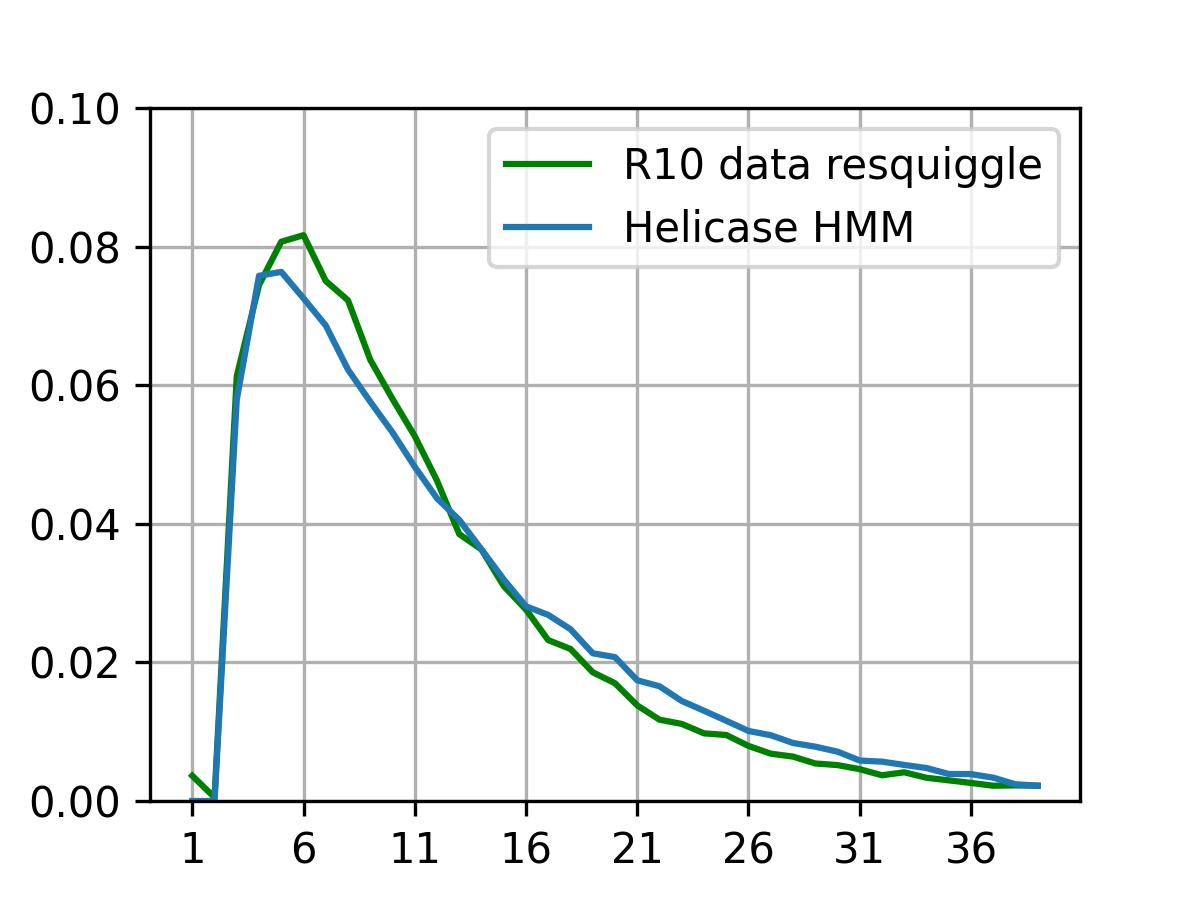}
  \caption{Duration histograms estimated from re-squiggle the R10 dataset (green) and simulated with the Helicase HMM inner-state transition (blue).}
  % JJ: This figure is not readable. I also do not understand why we need both this figure and Fig. 4
  \label{duration}
\end{figure}

We assume a first-order Markov chain for the $k$-mer sequence, where the outer states follow the one-base transitions rule that only allows a transition between two consecutive $k$-mers that overlap with $k-1$ bases. In addition, we assume a prior distribution $P(K)$ over $K\in \mathbb{K}$. Both the transition probability $P(K_{m+1}|K_{m})$ and the prior $P(K)$ can be estimated from a reference genome of the subject species or set to a uniform distribution for an unknown species. 

For the inner states, we assume they form a finite first-order Markov chain under the dominance of a $k$-mer outer state. At the beginning of an outer state's dominance, we draw from the set of inner states with a prior $P(H)$. The detailed transition mechanism of the inner states is illustrated in Fig.~\ref{HelicaseHMM}(b). These inner state transition probabilities are chosen to match the expected duration $L_m$, estimated from the training data. To validate this idea, we also plotted in Fig.~\ref{duration} the histograms of the duration $L_m$ per $k$-mer in R10 data
% JJ: We need to explain how the segmentation of re-squiggling was performed for this
and of the simulated duration based on the inner state transition probabilities. Proper duration estimation can be important for identifying the correct length of the homopolymers, illustrated in Fig.~\ref{signals}(b), where the current measurements are nearly identical from base to base and thus uninformative for the transitions. 

Finally, we assume dependency on both inner and outer states for a comprehensive feature space for the observation. Therefore, all inner states are emission states, each with independent observation space. Consequently, given a specific sequence of inner states, the ion current sequence can be re-indexed according to the inner states index as $\mathbf{X}=(X_1,...,X_N)=(X_{(1,1)}, ..., X_{(1,L_1)},X_{(2,1)},...,X_{(M,L_M)})$, such that $\sum^M_{m=1} L_m = N$. Similarly, the inner state sequence can be re-indexed as $(H_{(1,1)}, ..., H_{(1,L_1)},...,H_{(M,L_M)})=(H_1,...,H_N)$. We use a Gaussian distribution to characterize the specific state status, i.e., $P(X|K,H)\sim \mathcal{N}(\mu_{H,K},\sigma_{H,K})$. This necessitates over five million Gaussian distributions to characterize the ion currents comprehensively.

\begin{figure*}[!htp]
  \centering
  \includegraphics[width=0.9\textwidth]{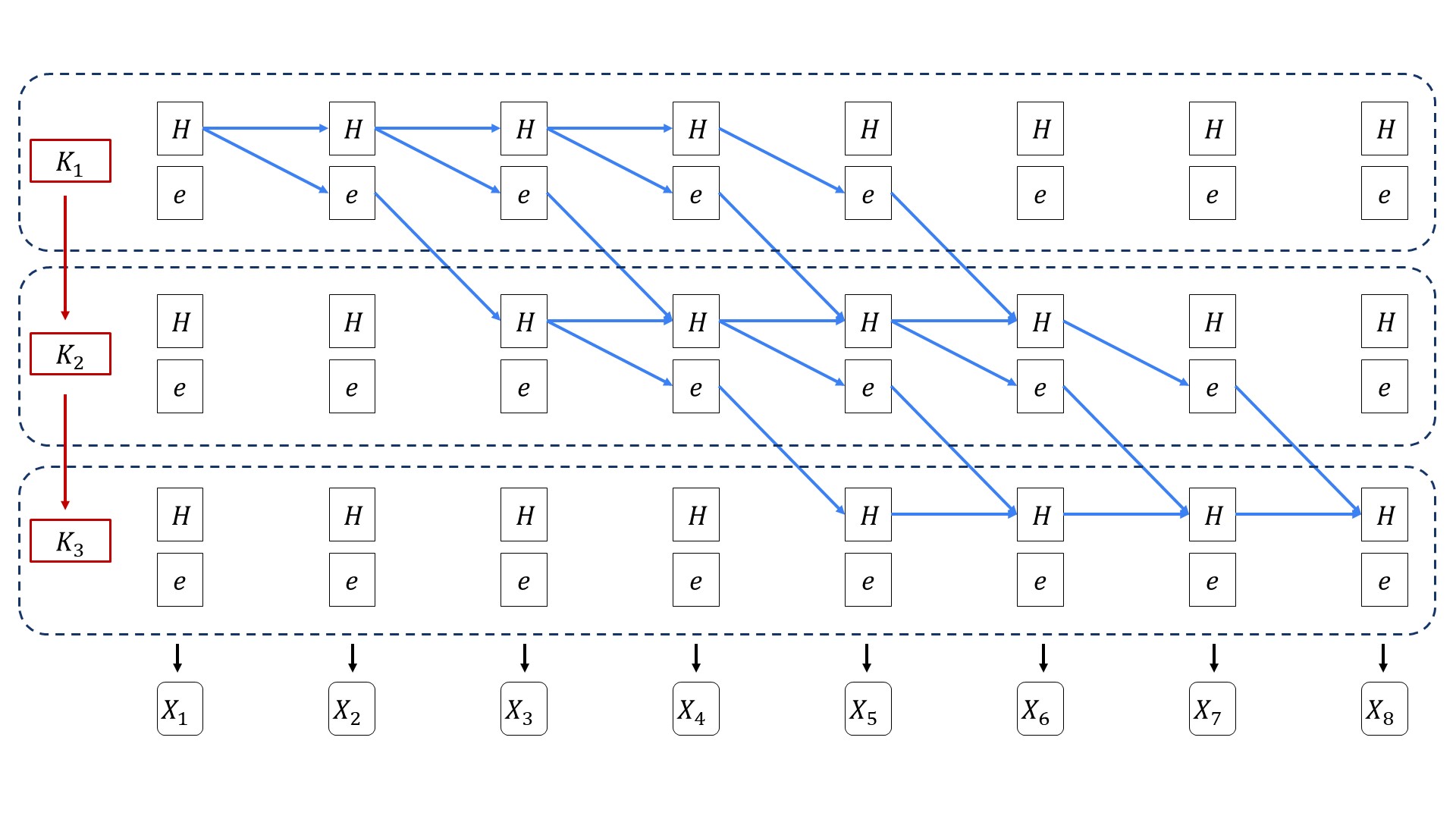}
  \caption{Illustration of the clamped-phase trellis folded according to a given $k$-mer sequence $(K_1,K_2,K_3)$. The helicase states are clamped into one box and separated from the end state for notational convenience. The arrows mark all possible transitions allowed with the Helicase HMM. Each path consists of arrows from left to right, indicating a segmentation of events.}
  \label{DBN}
\end{figure*}

\subsection{Algorithm and complexity}

In principle, the Helicase HMM, as a specific instance of the class of Hierarchical HMMs, can be inferred with linear complexity in the sequence length when the HMM is represented as a DBN \cite{linearHHMM}. However, practical implementation encounters challenges, particularly when dealing with a $10$-mer outer state space of size $4^{10}=1048576$, rendering algorithms operating on the entire state space computationally infeasible. For instance, computing the data likelihood $P(\mathbf{X})$ with $5$ inner states and an observation sequence length $N$ entails a complexity of $\mathcal{O}(5N4^{k})$
% JJ: We need to fix all Big-0 expressions that use fixed numbers
, surpassing the computational capabilities of commonly available processing units.

Hence, we pursue feasible alternatives, such as the joint likelihood $p(\mathbf{K},\mathbf{X})$ with reduced complexity of $\mathcal{O}(5NM)$ by marginalizing $p(\mathbf{K},\mathbf{H},\mathbf{X})$ over the set of all possible inner state sequences $\mathbf{H}$ of the length $N$ corresponding to the $k$-mer outer state sequence $\mathbf{K}$, denoted by $\mathbb{B}_N(\mathbf{K})$, as follows:
\begin{equation*}
\label{marginalH}
    p(\mathbf{K},\mathbf{X}) = \sum_{\mathbf{H}\in\mathbb{B}_N(\mathbf{K})} p(\mathbf{K},\mathbf{H},\mathbf{X})\text{ .}
\end{equation*}

As detailed in the preceding section and depicted in Fig. \ref{DBN}, the set $\mathbb{B}_N(\mathbf{K})$ can be fully represented by a rearranged DBN. Consequently, marginalization over the set can be accomplished by executing the forward-backward algorithm on the rearranged DBN.

The probability $P(\mathbf{K},\mathbf{H},\mathbf{X})$ given an outer state sequence $\mathbf{K}$, the ion currents $\mathbf{X}$, and an arbitrary inner state sequence $\mathbf{H}$, can be computed using eq.\eqref{eq:jointlike}. For notational convenience we assume that $P(K_1|K_0)$ and $P(H_{(m,1)}|H_{(m,0)})$ are the prior $P(K)$ and $P(H)$, respectively.
% what about duration?

\subsubsection{Training Algorithms}

In the context of the Helicase HMM, a direct adaptation of Maximum Likelihood training methods commonly used for standard HMMs, such as the Baum-Welch algorithm, is impractical due to the considerable computational cost imposed by such massive state space. However, training the Helicase HMM by maximizing the joint likelihood, denoted as  $\argmax P_n(\mathbf{K},\mathbf{X})$, also known as the clamped-phase likelihood in \cite{riis1998hidden}, is feasible and equivalent to executing the Baum-Welch algorithm on the rearranged DBN. 

To elaborate, one can employ the forward-backwards algorithm on the rearranged DBN to infer $P_n(K_m=i,H_n=j|\mathbf{X})$ for $i\in\mathbb{K}$ and $j\in\mathbb{H}$ at time $n$. For example, in Fig. \ref{DBN}, the forward propagation is the same as computing $P(\mathbf{K},\mathbf{X})$ along the blue arrows, while the backward propagation follows the blue arrows in opposite directions.

Then, the Helicase HMM's observation probabilities can be learned through the adapted Baum-Welch algorithm that maximizes the clamped-phase likelihood. For each iteration in the Baum-Welch algorithm, the updated mean $\mu_{(i,j)}$ and the variance $\Sigma_{(i,j)}=\sigma^2_{(i,j)}$ of the Gaussian observation probability for a given $k$-mer state $K=i$ and inner state $H=j$ can be computed as follows:
%\begin{equation*}
%    \begin{aligned}
%        & \mu_{(i,j)}=\frac{\sum_n\sum_mP_n(K_m=i,H_n=j,\mathbf{X})X_n}{\sum_n\sum_mP_n(K_m=i,H_n=j,\mathbf{X})} \text{ ,} \\
%        & \Sigma_{(i,j)} = \frac{\sum_n\sum_mP_n(K_m=i,H_n=j,\mathbf{X})(X_n-\mu_{(i,j)})^2}{\sum_n\sum_mP_n(K_m=i,H_n=j,\mathbf{X})}\text{, }
%    \end{aligned}
%\end{equation*}
\begin{align}
\mu_{(i,j)}=&\frac{\sum_n\sum_mP_n(K_m=i,H_n=j,\mathbf{X})X_n}{\sum_n\sum_mP_n(K_m=i,H_n=j,\mathbf{X})} \, \text{, and} \nonumber \\
\Sigma_{(i,j)} =& \frac{\sum_n\sum_mP_n(K_m=i,H_n=j,\mathbf{X})(X_n-\mu_{(i,j)})^2}{\sum_n\sum_mP_n(K_m=i,H_n=j,\mathbf{X})} \, . \nonumber
\end{align}
Here, $\sum_n\sum_mP_n(K_m=i,H_n=j,\mathbf{X})$ represents the expected count of how often the particular state $(K=i,H=j)$ appears in the training dataset. Essentially, considering that the Baum-Welch algorithm is a special case of the Expectation-Maximization method, we confine the expectation step to the partial set of the state space available in the training dataset. Consequently, the quality and representativeness of the training data significantly impact the generalizability of the trained model.

%\begin{strip}
%\begin{align}
%   \label{eq:jointlike}
%    p(\mathbf{K},\mathbf{H},\mathbf{X}) = \prod^M_{m=1}p(K_m|K_{m-1})\prod^{L_m}_{l=1}p(H_{(m,l)}|H_{(m,l-1)})p(X_{(m,l)}|K_m,H_{(m,l)}) 
%\end{align}
%\end{strip}

\begin{figure*}[t]
\begin{equation} \label{eq:jointlike}
p(\mathbf{K},\mathbf{H},\mathbf{X}) = \prod^M_{m=1}p(K_m|K_{m-1})\prod^{L_m}_{l=1}p(H_{(m,l)}|H_{(m,l-1)})p(X_{(m,l)}|K_m,H_{(m,l)})
\end{equation}
\end{figure*}

\subsubsection{Decoding Algorithms}
The decoding problem, essential for the basecalling task in nanopore sequencing, involves determining the most probable $k$-mer sequence given ion current measurements, i.e., $\argmax_{\mathbf{K}} P(\mathbf{K},\mathbf{X})$. However, this problem becomes infeasible as the search space grows exponentially with sequence length. In practice, two approaches are commonly employed to approximate a solution to the decoding problem.

The first approach is to find the most likely paths, i.e., $\argmax_{(\mathbf{K},\mathbf{H})}P(\mathbf{K},\mathbf{H},\mathbf{X})$, typically achieved using the Viterbi algorithm. Subsequently, the $k$-mer sequence is extracted while discarding other states. Although popular for many HMM-based basecallers, applying this approach to Helicase HMMs with a $10$-mer state space is impractical due to the Viterbi algorithm's complexity of $\mathcal{O}(N5^24^{20})$.

The second approach employs a beam search, a forward-searching algorithm that mitigates computational costs by limiting the search space to a fixed-sized region. The standard Beam Search serves as a low-complexity alternative to the Viterbi algorithm and provides several approximations to the most likely path. 

In our previous work \cite{MBS}, we have extended the beam search to arbitrary Hierarchical HMMs and devised algorithms and heuristics to approximate the most likely $k$-mer outer states by marginalizing the inner states. This led to the development of two Marginalized Beam Search algorithms: the Greedy Marginalized Beam Search (MGBS) and the Local Focus Beam Search (LFBS). The MGBS has a complexity of $\mathcal{O}\big(NW5^2+5N4^{10}\log(4^{10}5)\big)$ where $W$ denotes the beam width, excels in handling long homopolymers. Conversely, the LFBS, with a complexity of $\mathcal{O}\big(N4^M5^2+N4^{10}\log(4^{10})\big)$ where $M$ represents the focus length,
% JJ: This needs to be rewritten because 'focus length' is a term that is unknown to the reader at this point of the text
and offers greater accuracy in distinguishing short sequences. Given the objective of analyzing long nanopore reads, we used the MGBS in the experiments detailed in the subsequent section.

%And decoding to find the most-likely $k$-mer sequence $\argmax P(\mathbf{K}|\mathbf{X})$ would require to search over a total number of $\sum_{n=1}^N 4^n$ possible $k$-mer sequence $\mathbf{K}$. For basecaller applications, the essential task is to find the most likely $k$-mer sequence given a sequence of measurements, i.e., to decode $\argmax p(\mathbf{K}|\mathbf{X})$. 

%initalization with bonito table

\section{Experiment}
\subsection{Experimental Design}
The data utilized in our experiment were obtained from the ONT's `Sequencing Genome in a Bottle samples' project. The sequencing was conducted with the Ligation Sequencing Kit V14 released in MinKNOW 23.04.05, followed by basecalling with the Dorado software and alignment to the reference genome using Minimap2. Due to resource constraints, we focused on a quarter of the reads from the GM12878(HG001) sample aligned to chromosome $3$ for training and chromosome $5$ for testing.

The experimentation was conducted using NVIDIA A100 GPUs provided via the Berzelius supercomputer cluster of the national academic infrastructure for supercomputing in Sweden (NAISS). Timing measurements are provided for the use of a single of these GPUs. Training involving one epoch of the dataset consuming $32$ GPU hours. Notably, each raw read was divided into chunks of length $4096$ for training and, subsequently, basecalled individually using Dorado. The resultant bases were then aligned to the original nucleotide sequences of the long read to establish ground truth labels. The Helicase HMM used for validation experiments was trained over $10$ epochs of the training data and can be found in the \textbf{GitHub repository}.

\subsection{Re-squiggle}
We first commenced the validation of the Helicase HMM with a \textit{re-squiggle} task, which refers to aligning the ion current measurements with the corresponding nucleotide sequence. Such re-squiggled alignment allows for low-level analysis of the data and can be used for methylation detection.
% JJ: Can we get a reference to a paper doing Methylation detection?
In our approach with the Helicase HMM model, the re-squiggling was achieved by taking the most-likely state $P_n(K,H|\mathbf{X})$ at each time $n$. Although a more precise alignment could be achieved using the Viterbi algorithm to find the most likely inner state sequence, i.e., $\argmax_{(\mathbf{H})}P(\mathbf{K},\mathbf{H},\mathbf{X})$, in this work we use the former as it sufficiently demonstrates the advantage of the Helicase HMM on the re-squiggle task.

For benchmarking purposes, we also applied the `precise signal to sequence mapping refinement' from the ONT Remora v$3.1.0$ toolbox. This process involves first rescaling the signal with a $k$-mer table provided by the ONT, which is claimed to be the expected value of each $9$-mer generated with proprietary internal software. Then the re-squiggling was carried out using dynamic programming. 

\begin{figure*}[!t]
\centering
\subfloat[]{\includegraphics[width=1\linewidth]{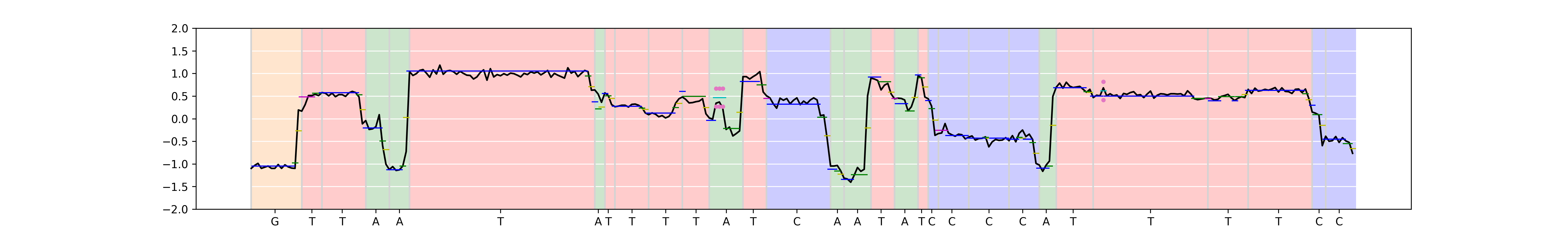}%
}
\hfil
\subfloat[]{\includegraphics[width=0.85\linewidth]{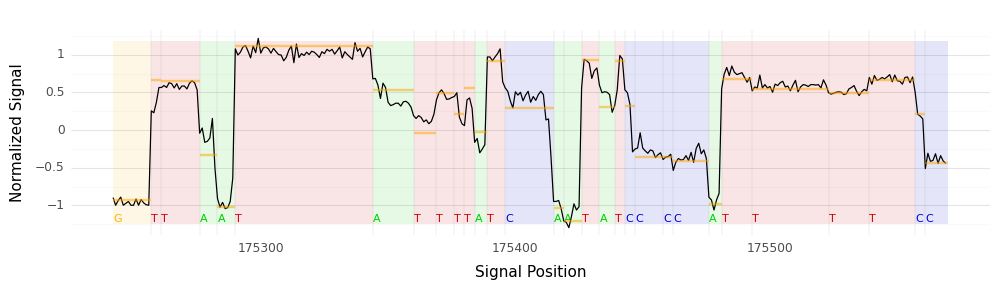}%
}
\caption{(a) Re-squiggle results obtained using the Helicase HMM are depicted in this figure. The signal levels corresponding to `step back' inner states are highlighted with pink dots, representing deviations of one standard deviation from the mean signal level on both sides. Additionally, different colored lines signify the mean signal level of each inner state for a specific $k$-mer. (b) Re-squiggle results obtained using ONT Remora are displayed in this figure. The $k$-mer table values are represented by yellow lines. A noticeable discrepancy between the two alignments occurs at the seventh to twelfth bases, specifically `ATTTTAT'.}
\label{resquiggle1}
\end{figure*}

\begin{figure*}[!t]
\centering
\subfloat[]{\includegraphics[width=1\linewidth]{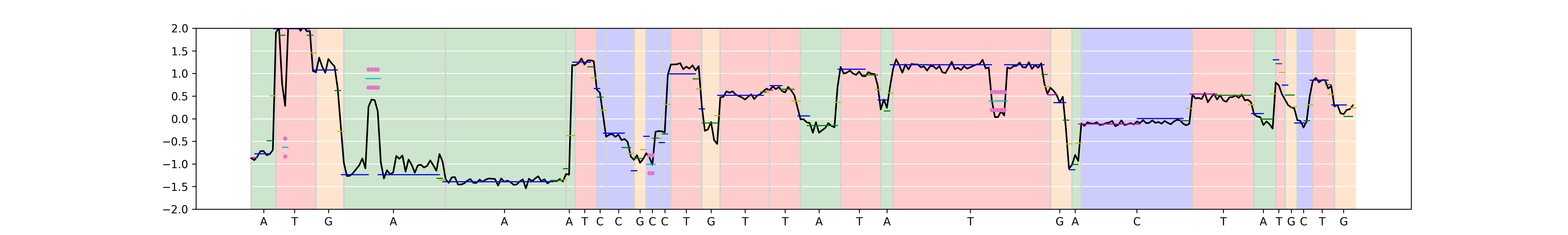}%
}
\hfil
\subfloat[]{\includegraphics[width=0.8\linewidth]{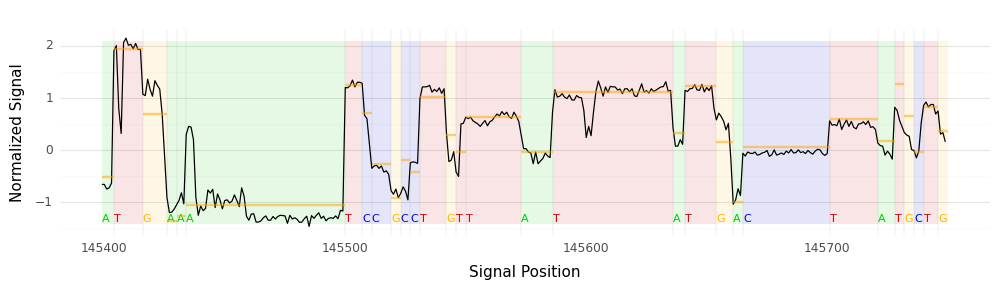}%
}
\caption{(a) Re-squiggle results with the Helicase HMM. (b) Re-squiggle results with ONT Remora. A discrepancy between the two alignments is observed at the third to fifth bases `AAA' and the eighteenth to twentieth bases `TAT'.}
% JJ: Can we redraw these figures to make them even more comparable? Maybe we can redraw the Remora alignment with your routine so all axes look the same.
\label{resquiggle2}
\end{figure*}

We present two instances of the re-squiggle results in Fig.~\ref{resquiggle1} and \ref{resquiggle2}, with the Helicase HMM results displayed on the top and the ONT Remora results at the bottom, adapted in a similar visual style. In the Helicase HMM results, the mean signal level of each inner state for a specific $k$-mer is depicted by different colored lines. For the `step back' inner states, we used pink dots to indicate signal levels at one standard deviation distance. Conversely, in the ONT Remora results, $k$-mer table values are marked as yellow lines.

In Fig.~\ref{resquiggle1}, the two results exhibit agreement in the alignment of most bases, for disagreement from the seventh to twelfth bases, i.e., `ATTTTTA'. Particularly at the twelfth base `A', the Helicase HMM identified signal fluctuations as consequences of `step backs'. However, the ONT Remora assigned the fluctuated signal to the previous `T' base, causing preceding mismatches of the $k$-mer tables and signals from the seventh base. Similarly, in Fig.~\ref{resquiggle2}, two disagreement areas occur at the third to fifth bases `AAA' and eighteenth to twentieth bases `TAT', where in both cases, the ONT Remora aligned the signal and the $k$-mer tables with large mismatches due to the signal fluctuation caused by the `step back'. In contrast, the Helicase HMM recognised both cases and managed to maintain a relatively close match between the expected signals and real signals.

\subsection{Basecalling Results}
%kmer vs acc
%with-without back step
In the second experiment, we validate the Helicase HMM with the basecalling task, using the MGBS decoder with beam width $512$. Trained on the data associated with Chromosome $3$, we reported the basecalling performance of the Helicase HMM basecaller associated with Chromosome $5$. Our Helicase HMM basecaller was implemented with CUDA and integrated as a customized operation within Python and TensorFlow. For benchmarking, we utilized ONT's Bonito basecaller, employing one of the pre-trained weights dna\_r10.4.1\_e8.2\_400bps\_fast@v4.2.0. Bonito, based on large neural networks, is implemented using Python and PyTorch.

Before basecalling, we re-normalize the data with a scale and a shift computed by linear regression between the raw currents and the predicted currents, which was obtained by first re-squiggling the raw currents using the Remora $k$-mer table and then assigning the value of each $k$-mer table to their aligned position. A neural network trained to predict such re-normalized currents can replace such a procedure.

The basecalling accuracy, or the identity score, is defined as the ratio of the number of matched bases to the total size of alignment areas and is calculated as follows:
\begin{equation*}
    \text{identity} = \frac{\text{matches}}{\text{matches}+\text{mismatches}+\text{insertion}+\text{deletions}}\; ,
\end{equation*}
where the alignment area begins at the first matched base and ends at the last matched base. The insertion, deletion, and substitution rates are similarly defined as their respective value divided by the alignment area. 

The performance metric also includes the basecalling speed, measured in terms of the number of bases output per second when running the basecaller on the Nvidia-V100 GPU with individually optimized batch size. It is important to note that only the time for the computational functions was recorded, as the Helicase HMM basecaller employs a non-optimized Input/Output handler in Python.

Table~\ref{basecalling} provides a comparison of the performance metrics of the two basecallers. Both basecallers demonstrate commendable accuracy in identifying bases within the sequence. Bonito Fast outperforms Helicase HMM in identity scores by $7.2\%$, highlighting the effectiveness of large neural networks. However, in terms of throughput, Helicase HMM showcases a significantly higher processing speed, with a throughput of approximately $12.556 \times 10^6$ bases per second, whereas Bonito Fast operates at a notably lower speed of $0.979 \times 10^6$ bases per second.

\begin{table}[!h]
\caption{Basecalling performances on Chromosome $5$\label{chr5}} {\begin{tabular}{@{}llllllll@{}}\toprule Basecaller &
identity & insert &delete& substitute & bases/second \\\midrule
HelicaseHMM & $0.843$ & $0.062$ & $0.032$ & $0.063$ & $12.556\mathrm{e}6$ \\
BonitoFast & $0.915$ & $0.025$ & $0.027$ & $0.032$ &$0.979\mathrm{e}6$ \\ \bottomrule
\end{tabular}}\label{basecalling}
\end{table}

\section{Conclusion}
In this study, we proposed and investigated the effectiveness of the Helicase HMM in nanopore sequencing applications inspired by pioneering studies on nanopore sequencing and motor protein research. Through experimentation on re-squiggling and basecalling tasks and comparison with established methods, we have gained valuable insights into the Helicase HMM's capabilities and performance.

Our findings demonstrate that the Helicase HMM offers significant advantages in re-squiggling, where it excels in accurately aligning ion current measurements with nucleotide sequences. By effectively recognizing and accommodating signal fluctuations, particularly those caused by `step backs,' the Helicase HMM outperforms existing methods, thus providing more precise alignments.

Furthermore, in the basecalling task, our results indicate that while ONT's Bonito Fast achieves higher identity scores, the Helicase HMM exhibits superior processing speed. Leveraging CUDA implementation and optimized batch sizes, the Helicase HMM demonstrates remarkable throughput, making it a compelling option for applications requiring rapid sequence analysis.

In conclusion, our study highlights the Helicase HMM as a promising approach for nanopore sequencing. Its ability to handle signal dynamics and its impressive processing speed position it as a valuable tool in the field of genomic research and clinical diagnostics. Moving forward, further optimizations and refinements to the Helicase HMM could enhance its performance, paving the way for advancements in nanopore sequencing technology.

\section*{Acknowledgments}
%The authors acknowledge support from Patrik Ståhl and Nayanika Bhalla, and the National Genomics Infrastructure (NGI) in Stockholm. 
This work was supported by the Swedish Research Council Research Environment Grant QuantumSense [VR 2018-06169]. The computations and data handling were enabled by the Berzelius resource provided by the Knut and Alice Wallenberg Foundation at the National Supercomputer Centre.

%\begin{thebibliography}{1}
\bibliographystyle{IEEEtran}
\bibliography{references}

\begin{thebibliography}{10}
\providecommand{\url}[1]{#1}
\csname url@samestyle\endcsname
\providecommand{\newblock}{\relax}
\providecommand{\bibinfo}[2]{#2}
\providecommand{\BIBentrySTDinterwordspacing}{\spaceskip=0pt\relax}
\providecommand{\BIBentryALTinterwordstretchfactor}{4}
\providecommand{\BIBentryALTinterwordspacing}{\spaceskip=\fontdimen2\font plus
\BIBentryALTinterwordstretchfactor\fontdimen3\font minus
  \fontdimen4\font\relax}
\providecommand{\BIBforeignlanguage}[2]{{%
\expandafter\ifx\csname l@#1\endcsname\relax
\typeout{** WARNING: IEEEtran.bst: No hyphenation pattern has been}%
\typeout{** loaded for the language `#1'. Using the pattern for}%
\typeout{** the default language instead.}%
\else
\language=\csname l@#1\endcsname
\fi
#2}}
\providecommand{\BIBdecl}{\relax}
\BIBdecl

\bibitem{ontyoutube}
\BIBentryALTinterwordspacing
C.~G. Brown, ``No thanks, i’ve already got one.'' 2016. [Online]. Available:
  \url{https://www.youtube.com/watch?v=nizGyutn6v4}
\BIBentrySTDinterwordspacing

\bibitem{baseRawller}
\BIBentryALTinterwordspacing
M.~Stoiber and J.~Brown, ``Basecrawller: Streaming nanopore basecalling
  directly from raw signal,'' \emph{bioRxiv}, 2017. [Online]. Available:
  \url{https://www.biorxiv.org/content/early/2017/05/01/133058}
\BIBentrySTDinterwordspacing

\bibitem{ONTdelivery2016}
M.~Jain, H.~E. Olsen, B.~Paten, and M.~Akeson, ``The {O}xford {N}anopore
  {M}in{ION}: {D}elivery of nanopore sequencing to the genomics community,''
  \emph{Genome Biology}, vol.~17, 12 2016.

\bibitem{analysiswithHMM}
J.~Schreiber and K.~Karplus, ``Analysis of nanopore data using hidden {M}arkov
  models,'' \emph{Bioinformatics}, vol.~31, pp. 1897--1903, 6 2015.

\bibitem{Nanocall}
\BIBentryALTinterwordspacing
M.~David, L.~J. Dursi, D.~Yao, P.~C. Boutros, and J.~T. Simpson, ``Nanocall: an
  open source basecaller for {O}xford {N}anopore sequencing data,''
  \emph{Oxford Bioinformatics}, 2017. [Online]. Available:
  \url{https://academic.oup.com/bioinformatics/article/33/1/49/2525680}
\BIBentrySTDinterwordspacing

\bibitem{HMMHDP}
A.~C. Rand, M.~Jain, J.~M. Eizenga, A.~Musselman-Brown, H.~E. Olsen, M.~Akeson,
  and B.~Paten, ``Mapping dna methylation with high-throughput nanopore
  sequencing,'' \emph{Nature Methods}, vol.~14, pp. 411--413, 2 2017.

\bibitem{Nanoraw}
\BIBentryALTinterwordspacing
T.~Szalay and J.~A. Golovchenko, ``De novo sequencing and variant calling with
  nanopores using poreseq,'' \emph{Nature Biotechnology}, vol.~33, pp.
  1087--1091, 10 2015. [Online]. Available:
  \url{https://www.nature.com/articles/nbt.3360}
\BIBentrySTDinterwordspacing

\bibitem{Nanopolish}
J.~T. Simpson, R.~E. Workman, P.~C. Zuzarte, M.~David, L.~J. Dursi, and
  W.~Timp, ``Detecting dna cytosine methylation using nanopore sequencing,''
  \emph{Nature Methods}, 2017.

\bibitem{Chiron}
H.~Teng, M.~D. Cao, M.~B. Hall, T.~Duarte, S.~Wang, and L.~J. Coin, ``Chiron:
  translating nanopore raw signal directly into nucleotide sequence using deep
  learning,'' \emph{GigaScience}, 2018.

\bibitem{2022human}
\BIBentryALTinterwordspacing
S.~Nurk, S.~Koren, A.~Rhie, M.~Rautiainen, A.~V. Bzikadze, A.~Mikheenko, M.~R.
  Vollger, N.~Altemose, L.~Uralsky, A.~Gershman, S.~Aganezov, S.~J. Hoyt,
  M.~Diekhans, G.~A. Logsdon, M.~Alonge, S.~E. Antonarakis, M.~Borchers, G.~G.
  Bouffard, S.~Y. Brooks, G.~V. Caldas, N.-C. Chen, H.~Cheng, C.-S. Chin,
  W.~Chow, L.~G. de~Lima, P.~C. Dishuck, R.~Durbin, T.~Dvorkina, I.~T. Fiddes,
  G.~Formenti, R.~S. Fulton, A.~Fungtammasan, E.~Garrison, P.~G.~S. Grady,
  T.~A. Graves-Lindsay, I.~M. Hall, N.~F. Hansen, G.~A. Hartley, M.~Haukness,
  K.~Howe, M.~W. Hunkapiller, C.~Jain, M.~Jain, E.~D. Jarvis, P.~Kerpedjiev,
  M.~Kirsche, M.~Kolmogorov, J.~Korlach, M.~Kremitzki, H.~Li, V.~V. Maduro,
  T.~Marschall, A.~M. McCartney, J.~McDaniel, D.~E. Miller, J.~C. Mullikin,
  E.~W. Myers, N.~D. Olson, B.~Paten, P.~Peluso, P.~A. Pevzner, D.~Porubsky,
  T.~Potapova, E.~I. Rogaev, J.~A. Rosenfeld, S.~L. Salzberg, V.~A. Schneider,
  F.~J. Sedlazeck, K.~Shafin, C.~J. Shew, A.~Shumate, Y.~Sims, A.~F.~A. Smit,
  D.~C. Soto, I.~Sović, J.~M. Storer, A.~Streets, B.~A. Sullivan,
  F.~Thibaud-Nissen, J.~Torrance, J.~Wagner, B.~P. Walenz, A.~Wenger, J.~M.~D.
  Wood, C.~Xiao, S.~M. Yan, A.~C. Young, S.~Zarate, U.~Surti, R.~C. McCoy,
  M.~Y. Dennis, I.~A. Alexandrov, J.~L. Gerton, R.~J. O’Neill, W.~Timp, J.~M.
  Zook, M.~C. Schatz, E.~E. Eichler, K.~H. Miga, and A.~M. Phillippy, ``The
  complete sequence of a human genome,'' \emph{Science}, vol. 376, no. 6588,
  pp. 44--53, 2022. [Online]. Available:
  \url{https://www.science.org/doi/abs/10.1126/science.abj6987}
\BIBentrySTDinterwordspacing

\bibitem{epigenetic2022}
\BIBentryALTinterwordspacing
A.~Gershman, M.~E.~G. Sauria, X.~Guitart, M.~R. Vollger, P.~W. Hook, S.~J.
  Hoyt, M.~Jain, A.~Shumate, R.~Razaghi, S.~Koren, N.~Altemose, G.~V. Caldas,
  G.~A. Logsdon, A.~Rhie, E.~E. Eichler, M.~C. Schatz, R.~J. O’Neill, A.~M.
  Phillippy, K.~H. Miga, and W.~Timp, ``Epigenetic patterns in a complete human
  genome,'' \emph{Science}, vol. 376, no. 6588, p. eabj5089, 2022. [Online].
  Available: \url{https://www.science.org/doi/abs/10.1126/science.abj5089}
\BIBentrySTDinterwordspacing

\bibitem{Lokatt}
X.~Xu, N.~Bhalla, P.~St{\r a}hl, and J.~Jald{\'e}n, ``Lokatt: a hybrid {DNA}
  nanopore basecaller with an explicit duration {H}idden {M}arkov model and a
  residual {LSTM} network,'' \emph{BMC Bioinformatics}, vol.~24, 12 2023.

\bibitem{ONTR10}
M.~Sereika, R.~H. Kirkegaard, S.~M. Karst, T.~Y. Michaelsen, E.~A. Sørensen,
  R.~D. Wollenberg, and M.~Albertsen, ``Oxford nanopore r10.4 long-read
  sequencing enables the generation of near-finished bacterial genomes from
  pure cultures and metagenomes without short-read or reference polishing,''
  \emph{Nature Methods}, vol.~19, pp. 823--826, 7 2022.

\bibitem{Bonito}
O.~N.~T. plc., ``Nanoporetech/{B}onito: A {P}ytorch basecaller for {O}xford
  {N}anopore reads.'' \url{https://github.com/nanoporetech/bonito}, Feb. 2020.

\bibitem{helicasedynamic}
J.~M. Craig, A.~H. Laszlo, H.~Brinkerhoff, I.~M. Derrington, M.~T. Noakes,
  I.~C. Nova, B.~I. Tickman, K.~Doering, N.~F.~D. Leeuw, and J.~H. Gundlach,
  ``Revealing dynamics of helicase translocation on single-stranded {DNA} using
  high-resolution nanopore tweezers,'' \emph{PNAS}, vol. 114, pp.
  11\,932--11\,937, 2017.

\bibitem{MBS}
X.~{Xu} and J.~{Jald{\'e}n}, ``Marginalized beam search algorithms for
  hierarchical {HMM}s,'' \emph{arXiv e-prints}, p. arXiv:2305.11752, May 2023.

\bibitem{linearHHMM}
\BIBentryALTinterwordspacing
K.~P. Murphy and M.~Paskin, ``Linear-time inference in hierarchical hmms,'' in
  \emph{Advances in Neural Information Processing Systems}, T.~Dietterich,
  S.~Becker, and Z.~Ghahramani, Eds., vol.~14.\hskip 1em plus 0.5em minus
  0.4em\relax MIT Press, 2001. [Online]. Available:
  \url{https://proceedings.neurips.cc/paper_files/paper/2001/file/aebf7782a3d445f43cf30ee2c0d84dee-Paper.pdf}
\BIBentrySTDinterwordspacing

\bibitem{riis1998hidden}
S.~Riis, ``\BIBforeignlanguage{English}{Hidden markov models and neural
  networks for speech recognition},'' Ph.D. dissertation, Technical University
  of Denmark, 1999.

\end{thebibliography}
%\end{thebibliography}

\section{Biography}
\begin{IEEEbiographynophoto}{Xuechun Xu}
received her MS degree in System Robotics and Control from KTH Royal Institute of Technology in 2018. She is now pursuing her PhD degree in the department of Information Science and Engineering, KTH. Her research interest lies in bioinformatics and sequential modeling.
\end{IEEEbiographynophoto}
\begin{IEEEbiographynophoto}{Joakim Jald\'{e}n}
received his Ph.D. in 2007 respectively from KTH Royal Institute of Technology, Stockholm, Sweden, where he is currently a professor of signal processing. His research interest includes signal processing for life science applications., including nano-pore sequencing, analysis of immuno-assays, and cell tracking in time-lapse microscopy.
\end{IEEEbiographynophoto}

\vfill

\end{document}